\documentclass[prd,tightenlines,twoside,secnumarabic,
               onecolumn,floatfix,nofootinbib,showpacs,11pt]{revtex4}

\usepackage[english]{babel}
\usepackage{amsmath,amssymb,bm,slashed}
\usepackage{graphicx}
\usepackage[sort&compress]{natbib}
\usepackage{xcolor}
\usepackage[normalem]{ulem}
\usepackage{hyperref}
\usepackage{cleveref}
\usepackage{subfigure} 
\definecolor{red}{rgb}{1.0, 0, 0}

\allowdisplaybreaks

\bibpunct{[}{]}{,}{n}{}{,}

\newcommand{\BR}{\text{BR}}

\newcommand{\parenbar}[1]{\overset{
            \raisebox{-0.15em}{\scalebox{.4}{\textbf{(}}}
            \raisebox{-0.3em}{{\hspace{.03em}--\hspace{.05em}}}
            \raisebox{-0.15em}{\scalebox{.4}{\textbf{)}}}} {#1}}

\newcommand{\im}{\textrm{Im}}
\newcommand{\re}{\textrm{Re}}

\begin{document}

\title{Flavor and CP violation in Higgs decays}
\author{Joachim Kopp$^{1,2}$}            \email[Email: ]{jkopp@uni-mainz.de}
\author{Marco Nardecchia$^3$}            \email[Email: ]{m.nardecchia@damtp.cam.ac.uk}
\affiliation{
             $^1$ Max Planck Institut f\"ur Kernphysik, Saupfercheckweg 1, 69117 Heidelberg, Germany \\
             $^2$ PRISMA Cluster of Excellence and Mainz Institute for Theoretical Physics,
               Johannes Gutenberg University, 55099 Mainz, Germany \\
             $^3$ DAMTP, University of Cambridge, Wilberforce Road, Cambridge CB3 0WA, United Kingdom}
\date{\today} 

\begin{abstract}
Flavor violating interactions of the Higgs boson are a generic feature of
models with extended electroweak symmetry breaking sectors.  Here, we
investigate CP violation in these interactions, which can arise from
interference of tree-level and 1-loop diagrams.
We compute the CP asymmetry in flavor violating Higgs
decays in an effective field theory with only one Higgs boson and in a general
Type-III Two Higgs Doublet Model (2HDM).  We find that large ($\sim \mathcal{O}(10 \% )$)
asymmetries are possible in the 2HDM if one of the extra Higgs bosons has a
mass similar to the Standard Model Higgs.  For the poorly constrained decay
modes $h \to \tau\mu$ and $h \to \tau e$, this implies that large lepton charge
asymmetries could be detectable at the LHC.  We quantify this by comparing the
sensitivity of the LHC to existing direct and indirect constraints.
Interestingly, detection prospects are best if Higgs mixing is relatively
small---a situation that is preferred by the current data.  Nevertheless,
CP violation in $h \to \tau\mu$ or $h \to \tau e$ will only be observable
if nonzero rates for these decay modes are measured very soon.
\end{abstract}

\begin{flushright}
\end{flushright}

\maketitle

\section{Introduction}
\label{sec:intro}

Precision measurements of the Higgs sector of elementary particles are becoming
one of the major topics in the physics program at the Large Hadron Collider
(LHC).  The main goal of these measurements is to search for deviations from
Standard Model (SM) expectations that would herald the existence of new physics
at the TeV scale, such as additional Higgs bosons, as
predicted for example in supersymmetry, secondary sources of electroweak
symmetry breaking, for instance due to strong dynamics, or non-standard
couplings of the Higgs boson due to higher-dimensional operators.

Many of these extensions of the SM predict the recenly discovered particle \cite{Chatrchyan:2012ufa,Aad:2012tfa} (hereafter refered to as the Higgs boson) to possess flavor
non-diagonal couplings (see for example~\cite{Bjorken:1977vt, McWilliams:1980kj,
Barr:1990vd, DiazCruz:1999xe, Casagrande:2008hr, Giudice:2008uua,
AguilarSaavedra:2009mx, Agashe:2009di, Goudelis:2011un, Kanemura:2005hr,
Blankenburg:2012ex, Craig:2012vj, Harnik:2012pb, Davidson:2012ds,
Arhrib:2012mg, Arhrib:2012ax, Dery:2013rta, Atwood:2013ica, Arroyo:2013kaa,
Celis:2013xja, Chen:2013qta, Greljo:2014dka, Gorbahn:2014sha} and references
therein).  Existing constraints on some of these couplings are
surprisingly weak, especially when couplings to third generation fermions are
involved as in the processes $h \to \mu\tau$, $h \to e \tau$, $t \to h c$ and
$t \to h u$.  A number of search strategies for flavor violating Higgs
couplings has been proposed~\cite{Craig:2012vj, Harnik:2012pb,
Davidson:2012ds,Aad:2012ypy,Bressler:2014jta} and first experimental searches
for top--charm--Higgs couplings have been carried out by
ATLAS~\cite{TheATLAScollaboration:2013nia} and CMS~\cite{CMS:2014qxa}.

Also CP violation in Higgs decays is an active topic of research, with the main
focus being on its effects on the polarization of the final state particles in $h
\to t \bar{t}$, $ h \to ZZ^*$, $h \to \gamma\gamma$ and $h \to
\tau\tau$~\cite{Dell'Aquila:1988fe, Dell'Aquila:1988rx, Grzadkowski:1995rx,
McKeen:2012av, Bishara:2013vya, Harnik:2013aja, Chen:2014gka, Brod:2013cka,
Shu:2013uua,Dolan:2014upa}.

Here we bring the two topics together by investigating Higgs decays that
violate flavor and CP.  In particular, we consider possible asymmetries between
the processes $h \to \ell^{i-} \ell^{j+}$ and $h \to \ell^{i+} \ell^{j-}$, as
parameterized by the observable
\begin{align}
  A_{CP}^{\ell^i \ell^j} \equiv \frac{\Gamma(h \to \ell^{i-} \ell^{j+}) - \Gamma(h \to \ell^{i+} \ell^{j-})}
                                 {\Gamma(h \to \ell^{i-} \ell^{j+}) + \Gamma(h \to \ell^{i+} \ell^{j-})} \,,
\end{align}
where $ \ell^i, \ell^j =  \{ e,\ \mu,\ \tau  \}  $ and $i \neq j$.  This observable offers perhaps the most
direct way of searching for CP violation in Higgs decays and does not require considering
any differential cross sections.  On the downside, a
measurement of $A_{CP}^{\ell^i \ell^j}$ requires large integrated luminosity due to
the smallness of (usually loop-induced) CP violating effects in general, and
due to the possible smallness of the decay rates $\Gamma(h \to \ell^{i \pm}
\ell^{j \mp})$ themselves.  The current 95\% CL upper limit on the branching
ratios $\BR(h \to \tau \mu)$ and $\BR(h \to \tau e)$ is 13\% from LHC
searches~\cite{Harnik:2012pb, Aad:2012mea}, while the indirect limit on $\BR(h
\to \mu e)$ is $2 \times 10^{-8}$~\cite{Harnik:2012pb}.\footnote{Here and in
the following, we denote by $\BR(h \to \ell^i \ell^j)$ the combined branching
ratio for the processes $h \to \ell^{i+} \ell^{j-}$ and $h \to \ell^{i-} \ell^{j+}$. When
referring to the branching ratio into only one of these CP-conjugate final
states, we use the notation $\BR(h \to \ell^{i+} \ell^{j-})$.} We will therefore not
consider the decay $h \to \mu e$ in our phenomenological analysis.

In sec.~\ref{sec:model}, we derive analytic expressions for $A_{CP}^{\ell^i \ell^j}$ in an
effective theory of CP violation in the Higgs sector induced by new particles
above the electroweak scale and in several classes of Two Higgs Doublet Models.
We will argue that these scenarios are very generic and encompass very large
classes of extensions of the SM.  We then constrain combined flavor and CP
violation in Higgs decays from low-energy observables in
sec.~\ref{sec:constraints}, and we estimate the sensitivity of the LHC in
sec.~\ref{sec:lhc}.  We summarize and conclude in
sec.~\ref{sec:conclusions}.

\section{Flavor and CP violation in the Higgs sector}
\label{sec:model}

\subsection{Low energy effective field theory with only one Higgs boson}
\label{sec:EFT}

\begin{figure}
  \begin{center}
    \includegraphics[width=14cm]{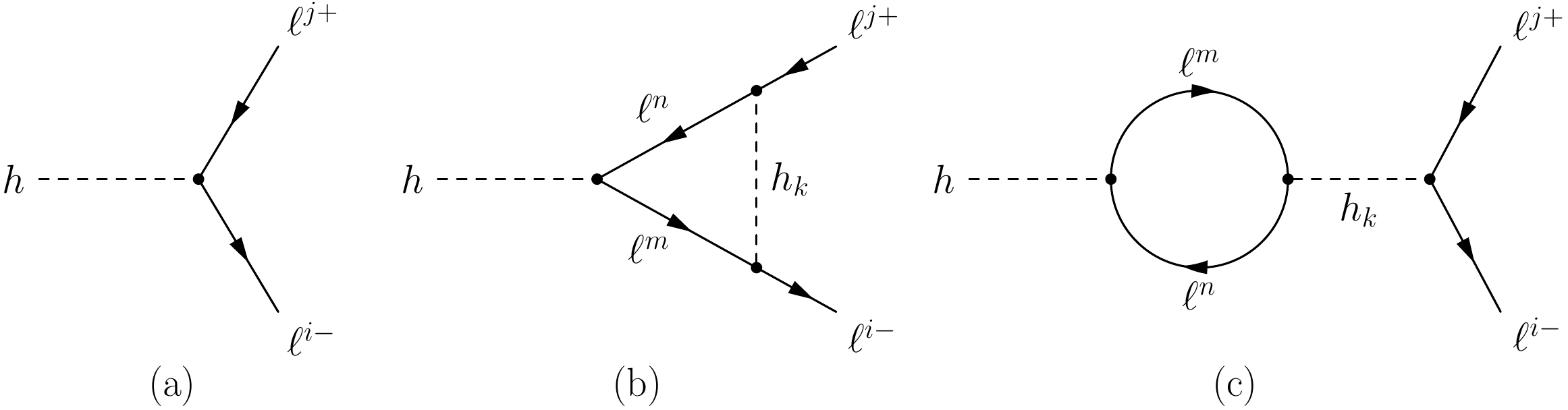}
  \end{center}
  \caption{Feynman diagrams contributing to flavor and CP violating Higgs boson decays
    $h \to \ell^{i-} \ell^{j+}$.  In the effective theory model
    (sec.~\ref{sec:EFT}), only one Higgs boson $h \equiv h_1$ exists and the bubble
    diagram (c) does not contribute to the CP asymmetry $A_{CP}^{\ell^i \ell^j}$. In the two
    Higgs doublet model (sec.~\ref{sec:2HDM}, there are three physical neutral
    Higgs mass eigenstates $h_1$, $h_2$, $h_3$, and all three diagrams
    contribute to $A_{CP}^{\ell^i \ell^j}$.}
  \label{fig:diagrams}
\end{figure}

We begin by considering the simplest low energy effective field theory (EFT) description
of flavor and CP violation in the Higgs sector,
\begin{align}
  {\cal L}_\text{EFT} &\supset -m_i \bar \ell_L^i \ell_R^i - Y_{ij}^h (\bar \ell_L^i \ell_R^j) h
                                + h.c. \,,
  \label{eq:L-EFT}
\end{align}
where $\ell^i$ are charged lepton fields in the mass basis, $h$ is the Higgs boson, and $Y_{ij}^h$ is a
general, complex $3 \times 3$ Yukawa matrix. Similar couplings can exist for quarks and
neutrinos, but we will here focus on the charged lepton sector, which is most easily accessible
at the LHC.

In the SM the couplings of the Higgs with the fermion fields are real and flavor diagonal while
several extensions of the SM predict $Y^h_{ij} \neq (m_i / v) \delta_{ij}$. 
Naturalness arguments suggest that the maximal size of the off diagonal elements should be related to the observed 
hierarchy of fermion masses. For example in \cite{Cheng:1987rs} in order to avoid tunings, relations like the following have to hold
\begin{equation}
\left| Y^h_{\mu \tau} Y^h_{\tau \mu} \right| \lesssim \frac{m_{\tau} m_{\mu}}{v^2} \,.
\end{equation}
Despite this general expectation one has to remark that the size of the flavor violating couplings is encoded in the details of the ultraviolet theory and in several explicit models larger flavor violating effects will be possible.
For this reason, in our approach we are not going to rely on any specific extension and we consider the couplings as free parameters.

The branching ratio for $h \to \ell^{i+} \ell^{j-}$ is given by
\begin{align}
  \BR(h \to \ell^{i+} \ell^{j-}) = \frac{\Gamma(h\to \ell^{i+} \ell^{j-})}
                                        {\Gamma(h\to \ell^{i+} \ell^{j-}) + \Gamma_\text{SM}} \,,
\end{align}
with
\begin{align}
  \Gamma(h \to \ell^{i+} \ell^{j-}) = \frac{m_h}{16 \pi} \big(|Y_{ji}^h|^2 + |Y_{ij}^h|^2 \big) \,
\end{align}
and with the SM Higgs width $\Gamma_\text{SM} = 4.1$~MeV for a 125~GeV Higgs
boson~\cite{Dittmaier:2011ti}.

The Lagrangian \eqref{eq:L-EFT} leads to non-zero $A_{CP}^{\ell^i \ell^j}$ through
interference of the first two diagrams shown in fig.~\ref{fig:diagrams}.  The
bubble diagram \eqref{eq:L-EFT} (c) exists, but does not contribute to
$A_{CP}^{\ell^i \ell^j}$. The tree level diagram is given by
\begin{align}
  i \mathcal{A}_\text{tree} = \bar\ell^i(p_i) \, i\big(Y_{ij}^h P_R
                                 + Y_{ji}^{h*} P_L \big) \, \ell^j (p_j) \,,
  \label{eq:Atree}
\end{align}
while the expression for the triangle diagram is
\begin{align}
  i \mathcal{A}_\text{triangle} &= \int\frac{d^4q}{(2\pi)^4}
    \bar\ell^i(p_i) \, i (Y_{im}^h P_R + Y_{mi}^{h*} P_L)
      \, \frac{i (\slashed{q} + \slashed{p_i} + m_m)}{(q + p_i)^2 - m_m^2}
      \, i (Y_{mn}^h P_R + Y_{nm}^{h*} P_L)  \nonumber\\
    &\hspace{6.5cm}
      \frac{i (\slashed{q} - \slashed{p_j} + m_n)}{(q - p_j)^2 - m_n^2}
      \, i (Y_{nj}^h P_R + Y_{jn}^{h*} P_L) \, \ell^j(p_j) \,.
  \label{eq:Atriangle}
\end{align}
Here, $p_i$, $p_j$ are the 4-momenta of the final state leptons, $m_i$ is the
mass of $\ell^i$, and $P_L = (1 - \gamma^5)/2$, $P_R = (1 + \gamma^5)/2$ are
the chirality projection operators.

From eqs.~\eqref{eq:Atree} and \eqref{eq:Atriangle}, we can compute
$A_{CP}^{\ell^i \ell^j}$ and find for the phenomenologically most interesting
case where $\ell^i=\mu$ and $\ell^j = \tau$
\begin{align}
  A_{CP}^{\mu\tau} &= 
    \frac{1 - \log{2}}{8 \pi} \, 
    \frac{ \im \left[ Y^h_{\tau \tau} \left( Y^{h}_{e \mu} Y^{h*}_{e \tau} Y^{h*}_{\mu \tau}
         - Y^{h}_{\mu e} Y^{h*}_{\tau e}  Y^{h*}_{\tau \mu} \right) \right]}
         {\left| Y^h_{\mu \tau} \right|^2 + \left| Y^h_{\tau \mu} \right|^2} \nonumber\\
  &\quad + \frac{1}{8 \pi} \, \frac{m^2_{\tau}}{m^2_h} 
    \frac{\left| Y^h_{\mu \tau} \right|^2 - \left| Y^h_{\tau \mu} \right|^2}
         {\left| Y^h_{\mu \tau} \right|^2 + \left| Y^h_{\tau \mu} \right|^2}
    \im \left[ (Y^h_{\tau \tau})^2 \right] \,.  
  \label{eq:ACP-EFT}
\end{align}
Here, we have neglected terms proportional to $m_\mu$, $m_e$, $\left| Y^h_{\mu\mu} \right|$,
$\left| Y^h_{ee} \right|$ as well as terms suppressed by more than one of the small
quantities $m_\tau^2 / m_h^2$, $|Y^h_{e\mu}|$ and $|Y^h_{\mu e}|$.
An analogous expression for $A_{CP}^{e\tau}$ is obtained by replacing
$\mu \leftrightarrow e$ in eq.~\eqref{eq:ACP-EFT}.

We see that, if only two lepton families (here $\tau$ and $\mu$) participate in
flavor changing Higgs couplings, $A_{CP}^{\mu \tau}$ is suppressed by
the loop factor $1/(8\pi)$ and by a factor $m_\tau^2 / m_h^2$.  When all three
lepton generations experience flavor changing Higgs couplings,
$A_{CP}^{\mu \tau}$ receives additional contributions that do not depend on
lepton masses, but are proportional to a product of three flavor violating
Yukawa couplings involving all three flavor combinations $e\mu$, $e\tau$ and
$\mu\tau$.  Since $(|Y_{\mu e}|^2 + |Y_{e\mu}|^2)^{1/2}$ is constrained to be smaller
than $3.6 \times 10^{-6}$ by searches for $\mu \to e \gamma$, $\mu \to e$
conversion in nuclei, and $\mu \to 3e$~\cite{Blankenburg:2012ex,
Harnik:2012pb}, these products are far too small for this source of CP
violation to be observable at the LHC.

We conclude that CP violation in flavor changing Higgs couplings is not
accessible at the LHC when the additional degrees of freedom responsible for
its generation are so heavy that they can be integrated out. We will therefore
now consider scenarios in which additional Higgs bosons appear as dynamical
degrees of freedom at the LHC. We will show that, in this case, large CP
asymmetries are possible.

\subsection{A type-III Higgs doublet model}
\label{sec:2HDM}

If not one but two Higgs doublets exist in nature~\cite{Lee:1973iz} (for a recent
review see \cite{Branco:2011iw}) and have masses of order
100~GeV, the phenomenology of the Higgs sector becomes considerably richer than
in the SM.  We will here consider a
general ``type-III'' Two Higgs Doublet Model (2HDM), in which both Higgs doublets
$\Phi_1$ and $\Phi_2$ have Yukawa couplings to charged leptons.  We work in
the Georgi basis \cite{Georgi:1978ri}, in which only $\Phi_1$ acquires
a vev so that $\Phi_1$ and $\Phi_2$ can be decomposed according to
\begin{equation}
  \Phi_1 = \begin{pmatrix}
             G^+ \\
             \frac{1}{\sqrt{2}} (v + \eta_1 + i G^0)
           \end{pmatrix}
  \qquad\qquad
  \Phi_2 = \begin{pmatrix}
             H^+ \\
             \frac{1}{\sqrt{2}} (\eta_2 + i A)
           \end{pmatrix} \,.
\end{equation}
Here $\eta_1$ and $\eta_2$ are real scalar fields, $A$ is a real
pseudoscalar, $H^\pm$ are the charged Higgs bosons and $G^\pm$, $G^0$ are Goldstone
bosons.  The charged lepton Yukawa couplings are given by
\begin{equation}
  \mathcal{L} \supset -\frac{\sqrt{2} m_i}{v}
    \delta_{ij} \, \bar L_L^i \ell_R^j \Phi_1 - \sqrt{2} Y_{ij} \, \bar L_L^i \ell_R^j \Phi_2 + h.c. \,,
\end{equation}
where $L_L^i$ denotes the lepton doublets and $(\delta_{ij})$, $(Y_{ij})$
are Yukawa matrices.  After electroweak symmetry breaking, the Yukawa couplings
become
\begin{equation}
  \mathcal{L} \supset -\left( \frac{m_i}{v} \delta_{ij} \, \eta_1
                      + Y_{ij} \, \eta_2
                      + i Y_{ij} \, A \right) \bar \ell_L^i \ell_R^j 
\end{equation}
In the most general Two Higgs Doublet Model, $\eta_1$, $\eta_2$ and $A$ are not
identical to the physical mass eigenstates, which we denote by $h_1$, $h_2$ and
$h_3$.  The two sets of fields are related by an orthogonal transformation
\begin{equation}
  (\eta_1, \eta_2, A)^T = O \cdot \, (h_1, h_2, h_3)^T \,,
\end{equation}
with $O \in SO(3)$.  We denote by $h_1$ the lightest mass eigenstate, which
is usally assumed to approximately resemble the SM Higgs boson. (Occasionally,
we will use the notation $h$ and $h_1$ interchangeably for this physical Higgs state.)
In the physical basis, the Lagrangian becomes
\begin{align}
  \mathcal{L} &= -m_i \bar \ell_L^i \ell_R^i
                 - \sum_{r=1,2,3} Y^{h_r}_{ij} \, \bar \ell_L^i \ell_R^j \, h_r  + h.c.
  \label{eq:Lhr} \\
\intertext{with}
  Y^{h_r}_{ij} &= \frac{m_i \delta_{ij}}{v} O_{1 r} + Y_{ij} O_{2 r}
                  + i Y_{ij} O_{3 r}  \,.
  \label{eq:Yhr}
\end{align}

For an arbitrary scalar mixing matrix $O$, the CP asymmetry in $h_1 \to \mu\tau$ decays
is given by
\begin{align}
  A_{CP}^{\mu\tau} = A_{CP}^{\mu\tau, (0)}
                     + A_{CP}^{\mu\tau, (1)} \frac{m_\tau}{v} + \dots \,,
  \label{eq:ACP-2HDM-general}
\end{align}
where ``\dots'' stands for terms that are second order in $m_\tau/v$ or first order in $m_\mu/v$, $m_e/v$, $|Y_{\mu\mu}|$,
$|Y_{ee}|$, $|Y_{e\mu}|$ or $|Y_{\mu e}|$. Setting $Y_{\mu
\mu}$ and $Y_{ee}$ to zero is motivated by the observed SM-like nature of the
125~GeV Higgs boson,  which suggests that $|Y_{\mu \mu}\left( O_{21}+iO_{31} \right)|$ and
$|Y_{ee}\left( O_{21}+iO_{31} \right)|$ cannot be larger than $\text{few} \times 10^{-2}$.
$Y_{e\mu}$ and $Y_{\mu e}$ in turn are tightly
constrained by searches for lepton flavor violation in $\mu \to e \gamma$, $\mu
\to e$ conversion in nuclei and $\mu \to 3e$~\cite{Blankenburg:2012ex,
Harnik:2012pb} (see also sec.~\ref{sec:EFT}).  The zeroth and first order terms
$A_{CP}^{\mu\tau, (0)}$ and $A_{CP}^{\mu\tau, (1)}$ in
eq.~\eqref{eq:ACP-2HDM-general} are given by
\begin{align}
  A_{CP}^{\mu\tau,(0)} &= \sum_{\alpha=2,3} \frac{1}{4 \pi} 
                  \frac{\left|Y_{\tau \mu}\right|^2 - \left|Y_{\mu \tau}\right|^2 }
                       {\left|Y_{\tau \mu}\right|^2 + \left|Y_{\mu \tau}\right|^2 }
                  \Big( \left|Y_{\mu \tau} \right|^2 + \left|Y_{\tau \mu} \right|^2
                      + \left|Y_{\tau \tau} \right|^2 \Big)
                   R_{\alpha}
                  \bigg[ g \bigg( \frac{m^2_{h_1}}{m^2_{h_{\alpha}}}  \bigg) 
                       + \frac{ m_{h_1}^2}{m_{h_1}^2-m_{h_{\alpha}}^2} \bigg] \,,
  \label{eq:ACP0-2HDM} \\
  A_{CP}^{\mu\tau,(1)} &= \sum_{\alpha=2,3} \frac{1}{8 \pi} 
                  \frac{\left|Y_{\tau \mu}\right|^2 - \left|Y_{\mu \tau}\right|^2 }
                       {\left|Y_{\tau \mu}\right|^2 + \left|Y_{\mu \tau}\right|^2 }
                   \left|Y_{\tau \tau} \right|
                  \bigg[ R^V_{\alpha} \; g \bigg( \frac{m^2_{h_1}}{m^2_{h_{\alpha}}} \bigg)
                       + R^L_{\alpha} \; g \bigg( \frac{m^2_{h_1}}{m^2_{h_{\alpha}}} \bigg)  
                       + R^L_{\alpha} \; \frac{2 m_{h_1}^2}{m_{h_1}^2-m_{h_{\alpha}}^2} \bigg] \,,
  \label{eq:ACP1-2HDM}
\end{align}
with the loop function
\begin{align}
  g(x) = \frac{x-\log(1+x)}{x} \,
  \label{eq:g}
\end{align}
and the vectors
\begin{align}
  R_{\alpha}   &= \frac{\left(O_{3 \alpha} O_{21} - O_{2 \alpha} O_{31} \right)
                        \left(O_{2 \alpha} O_{21} +O_{3 \alpha} O_{31} \right)}
                       {O^2_{21} + O^2_{31}} \,, \label{eq:R} \\
  R^V_{\alpha} &= \frac{ O_{2 \alpha} O_{21} +O_{3 \alpha} O_{31} }
                       {O^2_{21} + O^2_{31}}
\left[\sin \theta_{\tau} \left( O_{11} O_{2 \alpha}-O_{1 \alpha} O_{21}  \right) + \cos \theta_{\tau} \left( O_{11} O_{3 \alpha} - O_{1 \alpha} O_{31}\right) \right]
 \,, \label{eq:RV} \\
  R^L_{\alpha} &= \frac{O_{3 \alpha} O_{21} -O_{2 \alpha} O_{31} }
                       {O^2_{21} + O^2_{31}} 
\left[\cos \theta_{\tau} \left(O_{11} O_{2 \alpha} + O_{1 \alpha} O_{21} \right) - \sin \theta_{\tau} \left(O_{11} O_{3 \alpha} + O_{1 \alpha} O_{31}\right) \right]
\,. \label{eq:RL}
\end{align}
The angle $\theta_\tau$ is defined according to $Y_{\tau \tau} \equiv \left|
Y_{\tau \tau}\right| e^{i \theta_\tau}$.
The analogous expressions for $h_1 \to e\tau$ are again obtained by simply
replacing $\mu \leftrightarrow e$ in the above expressions.  We see that in the
general 2HDM, $A_{CP}^{\mu\tau}$ and $A_{CP}^{e\tau}$ are unsuppressed except
by the loop factor $1/(4\pi)$. Therefore the CP asymmetries can easily be of
order 10\%. If $m_{h_2} \simeq m_{h_1}$ or $m_{h_3} \simeq m_{h_1}$, even larger
asymmetries are possible as the last term in square brackets in eqs.~\eqref{eq:ACP0-2HDM}
and \eqref{eq:ACP1-2HDM} becomes large. Note, however, that in this case, the expansion
in $m_\tau / v$ from eq.~\eqref{eq:ACP-2HDM-general} breaks down, so
our analytic expressions are no longer directly applicable.

Eqs.~\eqref{eq:R}, \eqref{eq:RV} and \eqref{eq:RL} can be
simplified if we parameterize the scalar mixing matrix $O$ in terms of two mixing
angles $\theta_{12}$ and $\theta_{13}$,
\begin{align}
  O = \begin{pmatrix}
         c_{13} & 0 & s_{13} \\
         0      & 1 & 0 \\
        -s_{13} & 0 & c_{13}
      \end{pmatrix}
      \begin{pmatrix}
         c_{12} & s_{21} & 0 \\
        -s_{21} & c_{12} & 0 \\
         0      & 0      & 1
      \end{pmatrix} \,,
  \label{eq:O}
\end{align}
with the definitions $c_{ij} \equiv \cos\theta_{ij}$ and $s_{ij} =
\sin\theta_{ij}$. A third mixing angle $\theta_{23}$, corresponding to rotations
about the 1-axis, is unphysical because it can always be absorbed into a redefinition
of the of second Higgs doublet $\Phi_2 \to e^{i \theta_{23}} \Phi_2$.
With the explicit parameterization \eqref{eq:O}, eq.~\eqref{eq:R} becomes
\begin{align}
  R = (0,\, -r,\, r)^T \qquad \text{with} \qquad
  r \equiv \frac{s_{12} c_{12} s_{13} c_{13}^2}{c_{12}^2 s_{13}^2 + s_{12}^2} \,.
\end{align}
We see that $R_2 = -R_3$, i.e.\ that the contributions from $h_2$ and $h_3$ to
the CP violating loop diagrams tend to cancel each other in the limit $m_{h_2}
\approx m_{h_3}$. The explicit expressions for $R^V$ and $R^L$ are more lengthy.

We now consider several special cases of the general 2HDM.

\subsubsection*{Case 1: Heavy $h_3$}

Let us first consider a scenario where one of the neutral Higgs mass eigenstates,
say $h_3$, is much heavier than the other two.  The low energy effective Lagrangian
for this scenario is (see also eqs.~\eqref{eq:Lhr} and \eqref{eq:Yhr})
\begin{align}
  {\cal L}_{h_1 h_2} &= -m_i \bar f_L^i f_R^i - Y_{ij}^{h_1} (\bar f_L^i f_R^j) h_1
                                              - Y_{ij}^{h_2} (\bar f_L^i f_R^j) h_2  + h.c. \,.
\end{align}
The CP asymmetry is given by
\begin{align}
  A_{CP}^{\mu\tau, \text{case 1}}
    = \frac{1}{8 \pi} \, \bigg[ g \bigg(\frac{m^2_{h_1}}{m^2_{h_2}}\bigg)
        \frac{\im \left[ \left( Y^{h_2}_{\tau \mu} Y^{h_1 *}_{\tau \mu}
                              - Y^{h_2}_{\mu \tau} Y^{h_1 *}_{\mu \tau} \right)
                  \big( \sum_{ij} Y^{h_2}_{ij} Y^{h_1 *}_{ij} \big)  \right]}
             {\left| Y^{h_1}_{\mu \tau} \right|^2 + \left| Y^{h_1}_{\tau \mu} \right|^2} \nonumber\\
    + 2 \frac{m_{h_1}^2}{m_{h_1}^2-m_{h_2}^2}
        \frac{\im \left[ Y^{h_2}_{\tau \mu} Y^{h_1 *}_{\tau \mu}
                       - Y^{h_2}_{\mu \tau} Y^{h_1 *}_{\mu \tau} \right]
              \re \big[ \sum Y^{h_2}_{ij} Y^{h_1 *}_{ij} \big]}
             {\left| Y^{h_1}_{\mu \tau} \right|^2 + \left| Y^{h_1}_{\tau \mu} \right|^2}  \bigg] \,,
  \label{eq:ACP-2HDM-1}
\end{align}
with the loop function $g(x)$ from eq.~\eqref{eq:g}. We have again neglected
diagrams involving electrons.  We see that even in this considerably
simplified version of the 2HDM, unsuppressed CP violation can occur.

\subsubsection*{Case 2: Small mixing angles in the scalar sector}

The observed Standard Model-like nature of the 125~GeV Higgs boson suggests
that its mixing with the components of a heavy Higgs doublet should be small.
This leads us to consider the limit $\theta_{12}$, $\theta_{13} \ll 1$.  The
scalar mixing matrix thus becomes
\begin{align}
 O \approx \begin{pmatrix}
             1           &  \theta_{12} & \theta_{13} \\
            -\theta_{12} &  1           & 0 \\
            -\theta_{13} &  0           & 1
           \end{pmatrix}
\end{align}
and the Yukawa couplings in the physical basis are
\begin{align}
  \mathcal{L} &\supset
    - Y_{\tau\mu} \, \bar{\tau}_L \mu_R \,
      \big[ (\theta_{12} + i \theta_{13}) h_1 + h_2 + i h_3 \big]
    - Y_{\tau\mu} \, \bar{\nu}_{\tau L} \mu_R H^+ + (\mu \leftrightarrow \tau) + h.c.
                              & \text{(flavor violating)}  \nonumber\\[0.2cm]
  &\quad
    - \sum_i \frac{m_i}{v} \bar\ell^i_L \ell^i_R \,
        (h_1 + \theta_{12} h_2 + i \theta_{13} h_3) + h.c. \,.
                              & \text{(flavor conserving)}
  \label{eq:L-2HDM-physical}
\end{align}
This leads to the rate for $h_1 \to \tau^\pm \mu^\mp$,
\begin{align}
  \Gamma(h_1 \to \tau^+ \mu^-) = \frac{m_{h_1}}{16 \pi}
    \big(|Y_{\mu\tau}|^2 + |Y_{\tau\mu}|^2 \big) \,
    \big( \theta_{12}^2 + \theta_{13}^2 \big) \,.
  \label{eq:Gamma-small-theta}
\end{align}
The CP asymmetry is again given by eqs.~\eqref{eq:ACP-2HDM-general}--\eqref{eq:ACP1-2HDM},
but with eqs.~\eqref{eq:R}--\eqref{eq:RL} simplified to
\begin{align}
  R   &= \frac{1}{\theta_{12}^2 + \theta_{13}^2}
         \big(0,\, -\theta_{12} \theta_{13},\, \theta_{12} \theta_{13} \big)^T \,, \\
  R^V &= \frac{1}{\theta_{12}^2 + \theta_{13}^2}
         \big(0,\, -\theta_{12} \sin\theta_\tau,\, -\theta_{13} \cos\theta_\tau )^T \,, \\
  R^L &= \frac{1}{\theta_{12}^2 + \theta_{13}^2}
         \big(0,\, \theta_{13} \cos\theta_\tau,\, \theta_{12} \sin\theta_\tau )^T \,.
  \label{eq:R-small-theta}
\end{align}

\subsubsection*{Case 3: No CP violation in the scalar sector}

We now analyze the case where the scalar sector
is CP conserving. In the Georgi basis this means that $A = h_3$ is a mass eigenstate while
$\eta_1$ and $\eta_2$ can have a mixing. This mixing should not be too large so
that the lightest mass eigenstate $h_1$ is mostly $\eta_1$-like and behaves
like the SM Higgs boson, in agreement with LHC measurements of Higgs couplings.
Nevertheless, $h_1$ is still allowed to have an $\eta_2$ admixture of order
20\% as this is the accuracy to which the couplings of the 125~GeV Higgs boson
have been measured~(see for instance \cite{ATLAS:2014-HiggsCombi}).
Thus, sizeable flavor changing Yukawa couplings are still allowed.
We are thus led to consider a scalar mixing matrix of the form given by
eq.~\eqref{eq:O}, with $\theta_{13} = 0$ and $\theta_{12} \ll 1$.
With these approximations, $A_{CP}^{\mu\tau,(0)}$ in eq.~\eqref{eq:ACP-2HDM-general}
vanishes, and the leading term in the CP asymmetry, generated by loops involving
$h_2 \approx \eta_2$ and  $h_3 = A$ is given by $A_{CP}^{\mu\tau,(1)}$:
\begin{align}
  A_{CP}^{\mu\tau,\text{case 3}}
         &= -\frac{1}{8 \pi} \, \frac{\left|Y_{\tau \mu}\right|^2 - \left|Y_{\mu \tau}\right|^2}
                                    {\left|Y_{\tau \mu}\right|^2 + \left|Y_{\mu \tau}\right|^2} \,
             \frac{1}{\theta_{12}} \, \frac{m_{\tau}}{v} \, \im(Y_{\tau \tau}) \,
             \left[ g \left( \frac{m^2_{h_1}}{m^2_{h_2}}\right) - g \left( \frac{m^2_{h_1}}{m^2_{h_3}} \right)
                           - \frac{2 m_{h_1}^2}{m_{h_1}^2-m_{h_3}^2} \right]  
  \label{eq:ACP-2HDM-2}
\end{align}
The loop function $g(x)$ is again given by eq.~\eqref{eq:g}.  Note that
eq.~\eqref{eq:ACP-2HDM-2} is again based on the approximation $m_\mu = m_e =
Y_{\mu\mu} = Y_{ee} = 0$.  Eq.~\eqref{eq:ACP-2HDM-2} shows that in the 2HDM
without CP violation in the scalar sector the CP asymmetry is always suppressed
by $m_{\tau}/v$ and by $|Y_{\tau \tau}|$.

\section{Direct and Indirect Constraints}
\label{sec:constraints}

Direct and indirect searches constrain the maximal allowed amount of CP and flavor
violation in the Higgs decays accessible at the LHC.  In this section we summarize
how existing bounds on various low energy and high energy observables constrain
$\BR(h \to \ell^i \ell^j) \times A^{\ell^i \ell^j}_{CP}$.

\subsection{Effective theory with only one Higgs boson}
\label{sec:constraints-eft}

Assuming the flavor conserving couplings of the Higgs to quarks and
gauge bosons to be at their SM values, we derive in this section the
relevant bounds on the Yukawa couplings in the lepton sector.

Let us start by considering constraints coming from direct searches. Since we
assume for simplicity that flavor violating new physics affects predominantly
the lepton sector, the production cross section $\sigma$ for the Higgs boson
is the same as in the SM. It is therefore straightforward to use existing
searches for $h \to \tau^+ \tau^-$ and $h \to \mu^+ \mu^-$ to set bounds on the
flavor diagonal couplings $Y^h_{\tau \tau}$ and $Y^h_{\mu \mu}$.  The recent
CMS analysis~\cite{Chatrchyan:2014nva} finds a signal cross section for $h \to
\tau^+ \tau^-$ equal to $0.78 \pm 0.27$ times the standard model prediction.
This means equivalently that $\left| Y^h_{\tau \tau} \right|^2 / \left|(Y^h_{\tau
\tau})_\text{SM} \right|^2$ has to lie within this range. (Here and in the
following, the index SM denotes the SM values of the model parameters and
observables.) In a similar way the value of $Y^h_{\mu \mu}$ is bounded from
the results in~\cite{CMS:2013aga}, where $\Gamma(h \to \mu^+ \mu^-)$ has been
constrained to be smaller than 7.4 times its SM value.

The most recent constraints on flavor \emph{violating} couplings to the Higgs
boson to leptons have been derived in~\cite{Harnik:2012pb} by recasting an
ATLAS search for $h \to \tau^+\tau^-$~\cite{Aad:2012mea} in 4.7~fb$^{-1}$ of
7~TeV LHC data.  They require $\BR(h \to \tau \ell) < 0.13$, or equivalently
$\sqrt{|Y^h_{\tau\ell}|^2 + |Y^h_{\ell\tau}|^2} < 0.011$,
where $\ell = e, \mu$.  We expect that a similar analysis including data on $h
\to \tau^+ \tau^-$ from the 8~TeV run of the LHC could increase the sensitivity
to the Yukawa couplings by about a factor 1.5.  A dedicated analysis
could do significantly better still (see \cite{Davidson:2012ds,Bressler:2014jta} and sec.~\ref{sec:lhc}).
Note that a simple recasting of the existing $h \to \tau\tau$ searches at
$\sqrt{s} = 8$~TeV is not as promising as it was for the 7~TeV data used
in~\cite{Harnik:2012pb}.  In the case of the latest ATLAS
search~\cite{ATLAS:2013}, the reason is the usage of a boosted decision tree
which has been trained on SM $h \to \tau\tau$ decays and is therefore expected
to be less sensitive to other decay modes, in particular $h \to \tau\mu$ and $h
\to \tau e$.  The latest CMS search for $h \to \tau\tau$ is cut-based, but
employs a maximum likelihood method to determine the most likely value of the
Higgs mass on an event-by-event basis in spite of the incomplete kinematic
information.  This method is based on the assumption that any muon or electron
in the event originates from a $\tau$ decay and is thus accompanied by two
neutrinos.  Since this is not the case for $h \to \tau\mu$ and $h \to \tau e$
events, we expect the Higgs mass reconstruction to be very poor for the flavor
violating decay channels, leading to significant smearing of our signal and a
corresponding loss of sensitivity.

The direct bounds on the flavor-diagonal and flavor-off-diagonal Yukawa couplings
are summarized in the upper part of table~\ref{tab:leptons}. In the lower part,
we also show indirect constraints from the radiative decays
$\ell_i \to \ell_j + \gamma$~\cite{Blankenburg:2012ex,Harnik:2012pb}.
Other indirect observables like the electric and magnetic moments of the electron
and the muon give weaker bounds. (The electric dipole moment of the electron
leads, however, to a strong constraint on $\im(Y_{e\tau} Y_{\tau e}$.)
A more detailed discussion can be found, for example, in \cite{Blankenburg:2012ex,
Harnik:2012pb, Dery:2013rta}.

\begin{table}
  \centering
  \begin{ruledtabular}
  \begin{tabular}{@{\qquad}lcccr@{\qquad}}
    Channel               & Coupling                                       & Bound on coupling     & Bound on BR           & C.L. \\ \hline
    $h \to \tau^+ \tau^-$ & $|Y^h_{\tau \tau}|$                            & $8.3 \times 10^{-3}$  & 0.083                 & 95\% \\
    $h \to \mu^+ \mu^-$   & $|Y^h_{\mu \mu}|$                              & $1.1\times 10^{-3}$   & $1.6 \times 10^{-3}$  & 95\% \\
    $h \to \tau \mu$      & $\sqrt{|Y^h_{\tau\mu}|^2 + |Y^h_{\mu\tau}|^2}$ & 0.011                 & 0.13                  & 95\% \\
    $h \to \tau e$        & $\sqrt{|Y^h_{\tau e}|^2 + |Y^h_{e \tau}|^2}$   & 0.011                 & 0.13                  & 95\% \\
    \hline
    $\mu \to e \gamma$    & $\sqrt{|Y^h_{\mu e}|^2 + |Y^h_{e \mu}|^2}$     & $ 3.6 \times 10^{-6}$ & $2.4 \times 10^{-12}$ & 90\% \\ 
    $\mu \to e \gamma$    & $\big(|Y^h_{\tau\mu}Y^h_{\tau e}|^2 + |Y^h_{\mu\tau}Y^h_{e \tau }|^2\big)^{1/4}$
                                                                           & $ 3.4 \times 10^{-4}$ & $2.4 \times 10^{-12}$ & 90\% \\
    $\tau \to e\gamma$    & $\sqrt{|Y^h_{\tau e}|^2 + |Y^h_{e \tau}|^2}$   & $0.014$               & $3.3 \times 10^{-8}$  & 90\% \\
    $\tau\to \mu\gamma$   & $\sqrt{|Y^h_{\tau\mu}|^2 + |Y^h_{\mu\tau}|^2}$ & $0.016$               & $4.4 \times 10^{-8}$  & 90\%
  \end{tabular}
  \end{ruledtabular}
  \caption{Direct and indirect constraints on flavor conserving and flavor violating
  Yukawa couplings of the SM Higgs bosons in the effective theory eq.~\eqref{eq:L-EFT}.
  In the 2HDM, the constraints apply equivalently to $Y^{h_1}$ (see eq.~\eqref{eq:Yhr}).
  We have assumed that Higgs couplings to quarks and gauge bosons are
  unmodified compared to the SM.}
  \label{tab:leptons}
\end{table}

\subsection{Type-III Two Higgs doublet model}
\label{sec:constraints-2HDM}

Constraining the high dimensional parameter space of the general type-III Two Higgs
doublet model discussed in sec.~\ref{sec:2HDM} is a formidable task.  Here, our
goal is only to explore the region of parameter space where large CP violating
effects in flavor violating Higgs decays are possible and detectable at the LHC.  We therefore
simplify our analysis by assuming the mixing angles $\theta_{12}$ and
$\theta_{13}$ in the scalar sector to be small, and we set $\theta_{23}$ to
zero (cf.\ sec.~\ref{sec:2HDM}, case~2).  We will also assume that
$Y_{\tau\mu}$ and $Y_{\mu\tau}$ are the only nonzero element of the Yukawa
matrix $Y$.  This means that the second Higgs doublet couples to SM fermions
only through $Y_{\tau\mu}$ and $Y_{\mu\tau}$, and that the dominant decay modes
of the heavy Higgs mass eigenstates will be $h_2, h_3 \to \tau^\pm \mu^\mp$,
$H^\pm \to \mu^\pm \parenbar{\nu}_\tau$, $H^\pm \to \tau^\pm \parenbar\nu_\mu$ 

Decays to other combinations of SM fermions are possible due to Higgs mixing,
but since their rate is suppressed by the square of a small mixing angle, we
will neglect them. The heavy scalars can also decay through gauge interactions
as in the decay $H^{\pm} \to W^{\pm} h_i$. However, in the region of
parameter space where large $A^{\mu \tau}_{CP}$ can be observed at the LHC,
these decay channels are always subdominant.

The couplings of the SM-like Higgs mass eigenstate $h_1$ are constrained in the
same way as in sec.~\ref{sec:constraints-eft}. The bounds from table~\ref{tab:leptons}
translate into the limit
\begin{align}
  \sqrt{|Y_{\tau \mu}|^2 + |Y_{\mu \tau}|^2} \sqrt{\theta_{12}^2 + \theta_{13}^2} < 0.011 \,.
  \label{eq:Ytaumu-limit-2HDM}
\end{align}
We see from eqs.~\eqref{eq:ACP-2HDM-general}, \eqref{eq:ACP0-2HDM},
\eqref{eq:Gamma-small-theta} and \eqref{eq:R-small-theta}
that the largest observable CP violating effects,
as measured by
\begin{align}
  \Gamma(h_1 \to \tau^+\mu^-) \times A^{\mu \tau}_{CP}
    &\simeq -\frac{m_{h_1}}{64 \pi^2}
            \theta_{12} \theta_{13}
                  \big( |Y_{\tau \mu}|^2 - |Y_{\mu \tau}|^2 \big)
                  \big( |Y_{\mu \tau}|^2 + |Y_{\tau \mu}|^2 + |Y_{\tau \tau}|^2 \big) \nonumber\\
    &\quad\times
            \sum_{\alpha=2,3} (-1)^\alpha
                  \bigg[ g \bigg( \frac{m^2_{h_1}}{m^2_{h_{\alpha}}}  \bigg) 
                       + \frac{ m_{h_1}^2}{m_{h_1}^2-m_{h_{\alpha}}^2} \bigg] \,,
\end{align} 
are obtained if either $Y_{\mu\tau} = 0$, $Y_{\tau\mu} \neq 0$ or $Y_{\mu\tau} \neq 0$,
$Y_{\tau\mu} = 0$. Moreover, to obtain large CP violation,
the limit from eq.~\eqref{eq:Ytaumu-limit-2HDM} should be saturated,
$|Y_{\tau\tau}|$ should be of order $0.1 / \sqrt{\theta_{12}^2 + \theta_{13}^2}$ (larger values
are excluded by measurements of $\BR(h_1 \to \tau\tau)$)
and $\theta_{12} = \theta_{13}$. Finally, $h_2$ and $h_3$ should be very different
in mass since there is no CP violation if $m_{h_2} = m_{h_3}$.  Most interesting to us
is therefore the limit $m_{h_3} \gg m_{h_2} \approx m_{h_1}$.

Constraints on $h_2$ and $h_3$ from direct production are not important in the
small mixing angle limit since the production of the heavy Higgs mass eigenstates
is suppressed by $\theta_{12}^2$ or $\theta_{13}^2$.  If the dominant
decay mode of $h_2$ and $h_3$ is to $\tau + \mu$ as assumed here, conventional
searches for flavor conserving final states are suffering from an additional
mixing angle suppression in the flavor conserving branching ratios.
The strongest limits on $h_2$ and $h_3$ are therefore coming from indirect
searches, in particular $\tau \to \mu\gamma$.  We obtain these limits following
the procedure outlined in ref.~\cite{Harnik:2012pb}.  We match the full 2HDM
onto the effective Lagrangian
\begin{align}
  \mathcal{L} = c_L Q_{L\gamma} + c_R Q_{R\gamma} + h.c. \,,
\end{align}
with the operators
\begin{align}
  Q_{L\gamma, R\gamma} &= \frac{e}{8\pi^2} m_\tau
    \big(\bar\mu\,\sigma^{\alpha\beta} P_{L,R} \tau\big) F_{\alpha\beta} \,.
\end{align}
Here, $F_{\alpha\beta}$ is the electromagnetic field strength tensor.
The Wilson coefficients $c_L$, $c_R$ receive contributions from
one-loop diagrams involving neutral Higgs boson--charged lepton bubbles and
from two-loop diagrams containing top or $W$ loops.  For simplicity, we neglect
diagrams involving the charged Higgs bosons $H^\pm$, assuming they are sufficiently heavy.
The contributions of $h_1$ to $c_L$ and $c_R$ are given by the expressions
summarized in the appendix of~\cite{Harnik:2012pb}, with the modification that,
following eq.~\eqref{eq:L-2HDM-physical}, $Y_{\tau\mu}$ and $Y_{\mu\tau}$ are
replaced by $Y_{\tau\mu} (\theta_{12} + i \theta_{13})$ and $Y_{\mu\tau}
(\theta_{12} + i \theta_{13})$, respectively. Similarly, for the contribution
of diagrams containing $h_2$ ($h_3$), the flavor-diagonal Yukawa couplings as
well as the Higgs couplings to gauge bosons have to be multiplied by
$-\theta_{12}$ ($-\theta_{13}$).  For the $h_3$ contributions, moreover,
$Y_{\tau\mu}$ is replaced by $i Y_{\tau\mu}$ and $Y_{\mu\tau}$ by $i
Y_{\mu\tau}$.

A second set of indirect limits on 2HDMs arises from measurements of the
electric and magnetic dipole moments of the electron and muon.  If the only
non-negligible Yukawa couplings of the second Higgs doublet are $Y_{\tau\mu}$
and $Y_{\mu\tau}$, the one-loop contributions of the heavy Higgs bosons to the
electric (magnetic) dipole moment $d_\mu$ ($a_\mu$) of the muon are proportional to
$\Re(Y_{\tau\mu} Y_{\mu\tau})$ ($\Im(Y_{\tau\mu} Y_{\mu\tau})$). They are, however
\emph{not} suppressed by the mixing angles $\theta_{12}$ and $\theta_{13}$.
However, as we have seen above, large CP violation in $h_1 \to \tau\mu$ is only
possible if $Y_{\tau\mu}$ and $Y_{\mu\tau}$ are very different in magnitude.  In this
case, dipole moment constraints deteriorate rapidly and we will therefore not consider
them further here.

\begin{figure}
  \begin{center}
    \includegraphics[width=0.48\textwidth]{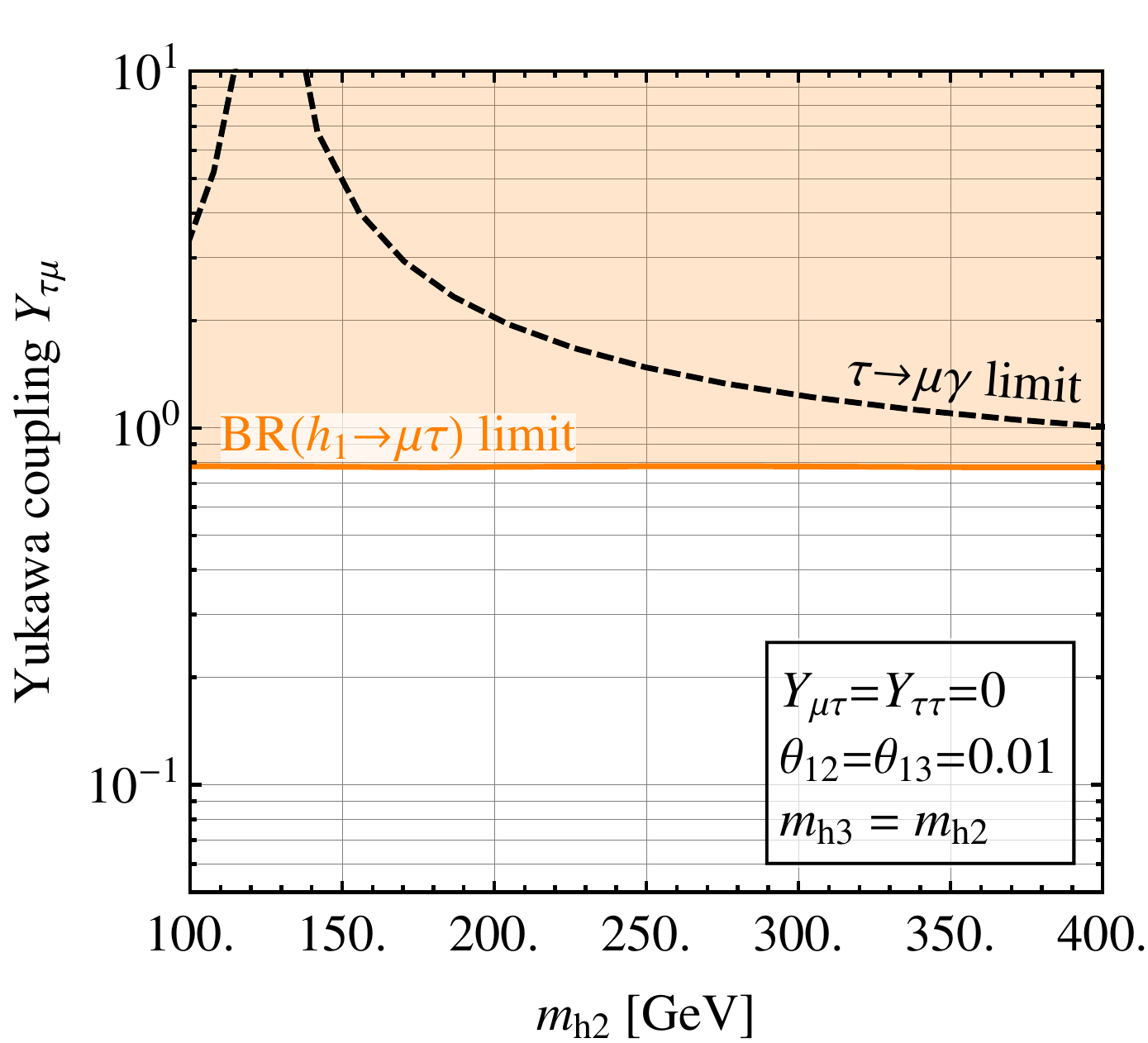} \hfill
    \includegraphics[width=0.48\textwidth]{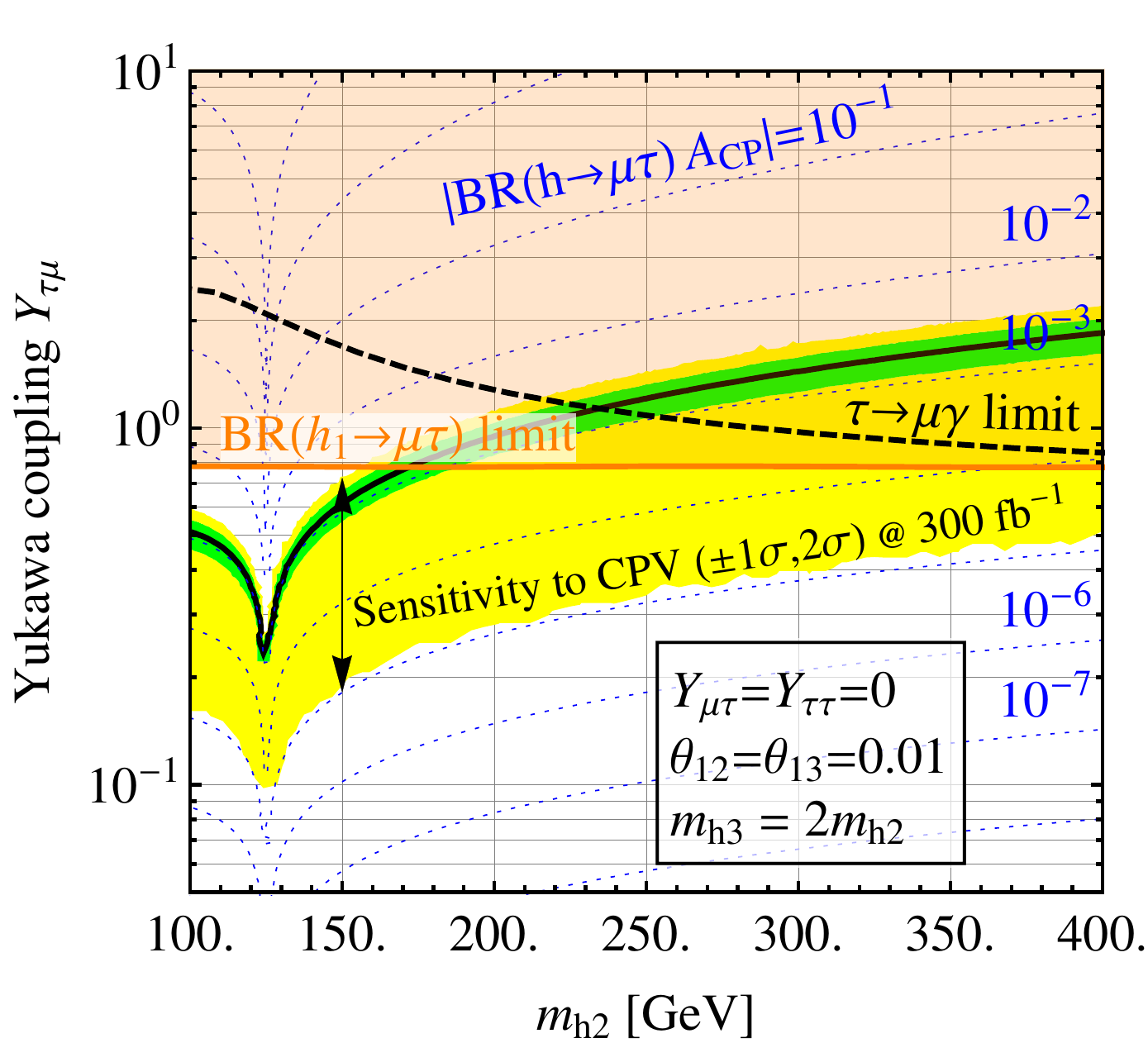}
  \end{center}
  \caption{
    Direct and indirect constraints on the flavor violating Yukawa couplings
    $Y_{\tau\mu}$ in the general type-III 2HDM.  We have assumed all other
    entries of the Yukawa matrix $Y$ (including in particular $Y_{\mu\tau}$) to
    vanish.  In the left panel, we have assumed $m_{h_2} = m_{h_3}$, a situation
    in which no CP violation is expected, while in the right panel, we consider
    a benchmark scenario with $m_{h_3} > m_{h_2}$.  We show the region excluded
    by the direct limit on $\BR(h_1 \to \tau\mu)$ from LHC
    data~\cite{Harnik:2012pb} (orange) together with indirect limits from $\tau
    \to \mu\gamma$ (black dashed).  In the right panel, the
    ``Brazilian band'' (black curve with green and yellow $\pm 1\sigma$ and
    $\pm 2\sigma$ bands) indicates the expected 95\% C.L.\ limit from a
    search for CP violation at the 13~TeV LHC with an integrated luminosity of
    300~fb$^{-1}$ (see sec.~\ref{sec:lhc} for details).  The regions above the
    band is approximately equal to the region in which evidence for CP
    violation can be found.  The blue dotted contours indicate constant values
    of the quantity $|\BR(h_1 \to \tau\mu) \times A_{CP}^{\mu\tau}|$, which is a
    measure for the observability of CP violation.  The largest CP violating
    effects are expected for $m_{h_2}$ similar to the mass of the SM-like Higgs
    boson. (Note that in this case, our plots are only approximate since the
    underlying analytic expansion from eq.~\eqref{eq:ACP-2HDM-general} breaks down.)}
  \label{fig:limit-mh2}
\end{figure}

\begin{figure}
  \begin{center}
    \includegraphics[width=0.48\textwidth]{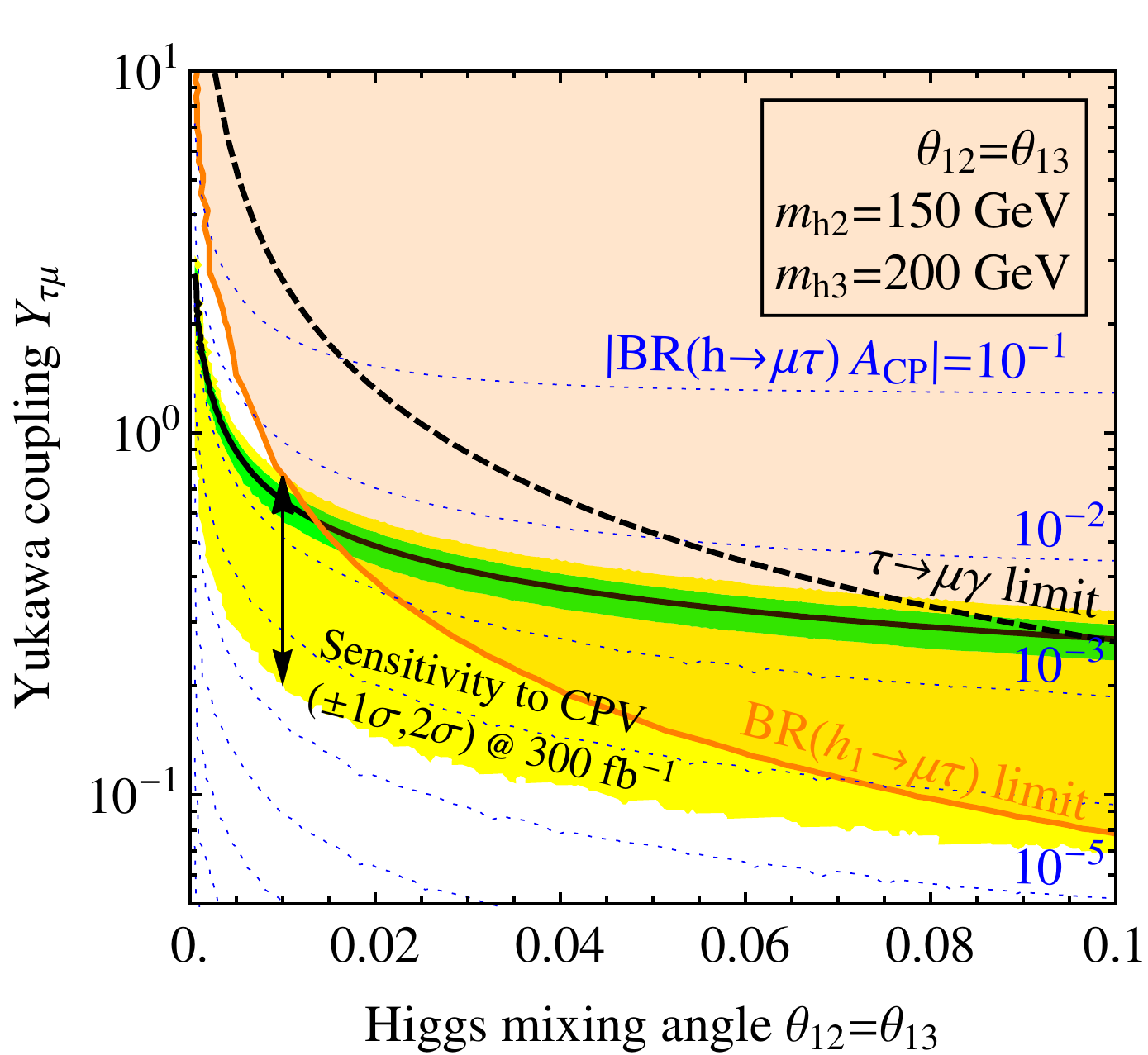} \hfill
    \includegraphics[width=0.48\textwidth]{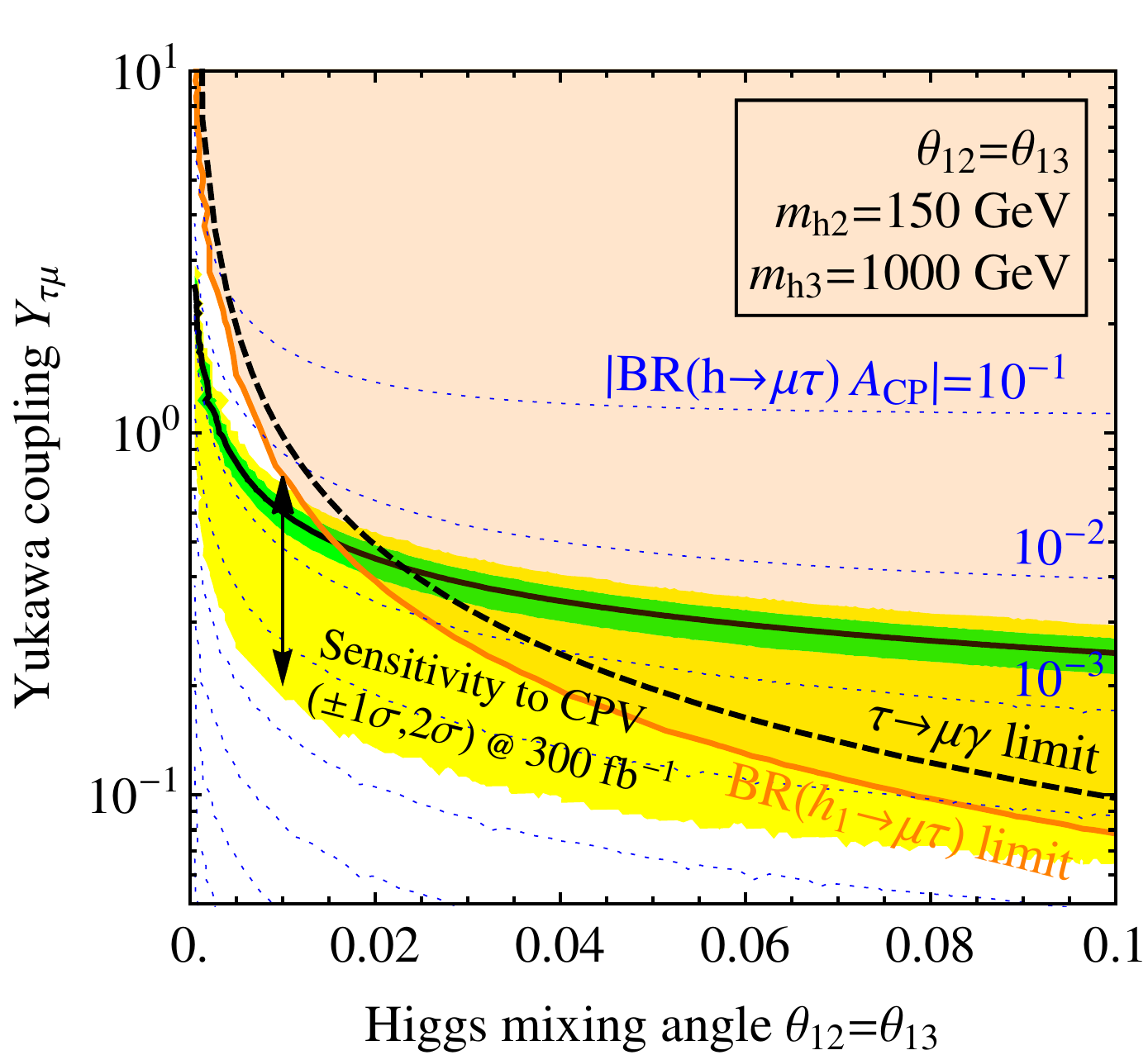}
  \end{center}
  \caption{Direct and indirect constraints on the flavor violating Yukawa couplings
    $Y_{\tau\mu}$ in the general \mbox{type-III} 2HDM for fixed Higgs boson masses
    $m_{h_2}$ and $m_{h_3}$, but varying Higgs mixing angles.  We have assumed
    all entries of the Yukawa matrix $Y$ other than $Y_{\tau\mu}$ to vanish.
    We show the region excluded by the direct limit on $\BR(h_1 \to \tau\mu)$
    from LHC data~\cite{Harnik:2012pb} (orange) together with indirect limits
    from $\tau \to \mu\gamma$ (black dashed).  The ``Brazilian bands'' (black
    curves with green and yellow $\pm 1\sigma$ and $\pm 2\sigma$ bands)
    indicate the expected 95\% C.L.\ limits from a search for CP violation at
    the 13~TeV LHC with an integrated luminosity of 300~fb$^{-1}$ (see
    sec.~\ref{sec:lhc} for details).  The regions above the bands are
    approximately equal to the regions in which evidence for CP violation can
    be found.  The blue dotted contours indicate constant values of the
    quantity $|\BR(h_1 \to \tau\mu) \times A_{CP}^{\mu\tau}|$, which is a measure for
    the observability of CP violation.  A search for CP violating effects is
    most promising if the mixing angles are small.}
  \label{fig:limit-theta}
\end{figure}

In figs.~\ref{fig:limit-mh2} and \ref{fig:limit-theta}, we compare the indirect
$\tau \to \mu\gamma$ constraints on the Yukawa couplings
(black dashed curves) and the direct constraint $\BR(h_1 \to \tau\mu) <
0.13$~\cite{Harnik:2012pb} (orange shaded region) to the expected $\BR(h_1 \to
\tau\mu) \, A_{CP}^{\mu\tau}$ (blue dotted contours).  The latter quantity is a
measure for the observability of CP violation in $h_1 \to \tau\mu$ decays.  We
also show the expected sensitivity of the LHC to CP violation
in $h_1 \to \tau\mu$ decays (see sec.~\ref{sec:lhc} for details).  For
illustration, we have here assumed that $Y_{\tau\mu}$ is the only nonzero
element of the Yukawa matrix $Y$ since we see from eqs.~\eqref{eq:ACP0-2HDM}
and \eqref{eq:ACP1-2HDM} that a large asymmetry between $|Y_{\tau\mu}|$ and
$|Y_{\mu\tau}|$ maximizes the CP asymmetry. Note that in the left panel of
fig.~\ref{fig:limit-mh2}, no CP violation is expected because we have assumed
$m_{h3} = m_{h2}$ there.  We see from figs.~\ref{fig:limit-mh2} and
\ref{fig:limit-theta} that the largest observable CP violation is expected when
$m_{h2} \sim m_{h1}$ and $m_{h3}$ much heavier.  Moreover, the Higgs mixing
angles $\theta_{12}$ and $\theta_{13}$ should be small---a situation that is
actually preferred by the current LHC data, which is very SM-like.

Finally, we comment on constraints on the charged Higgs bosons $H^\pm$ whose
quantum numbers are the same as those of a left-handed slepton in
supersymmetry. Therefore, limits on slepton masses from direct production at
the LHC can in principle be recast into limits on the charged Higgs boson mass
$m_{H^\pm}$. The ATLAS slepton search in 20.3~fb$^{-1}$ of 8~TeV
data~\cite{Aad:2014vma} constrains the mass of left-handed sleptons to be
$m_{\tilde\ell} \gtrsim 300$~GeV, assuming a simplified scenarios with
mass-degenerate left-handed selectrons and smuons, massless neutralinos, and
all other SUSY particles very heavy.  Comparing slepton pair production in this
simplified SUSY model to the production of $H^\pm$ of the same mass in our
2HDM, we note that $H^\pm$ production leads to about a factor of 8 fewer
events. The reason is that there are two new particles (selectron and smuon) in
the SUSY scenario, but only one new particle in the 2HDM. Moreover, $p p \to
H^+ H^-$ has a branching ratio to the dimuon + MET final state of only
25\%---the remainder of the events contains one or two tau leptons. Therefore,
the bound on $m_{H^\pm}$ is significantly weaker than the one on
$m_{\tilde\ell}$, so requiring the charged Higgs bosons to be heavier than
300~GeV is a very conservative assumption.

\section{Flavor and CP violating Higgs decays at the LHC}
\label{sec:lhc}

To investigate the sensitivity of future LHC searches to the CP asymmetry
$A_{CP}^{\mu\tau}$ in the decay $h \to \tau\mu$, we follow the strategy proposed
in~\cite{Davidson:2012ds}.  (Results for $h \to \tau e$ will be very similar.)
The search proposed there is sensitive to Higgs
boson production through gluon fusion, and is therefore expected to be more
sensitive than the alternative strategy proposed in~\cite{Harnik:2012pb}, which
is optimized for Higgs production through vector boson fusion.
We adapt the method outlined in~\cite{Davidson:2012ds} to a hadronic center of
mass energy $\sqrt{s} = 13$~TeV and an integrated luminosity of 300~fb$^{-1}$.
We normalize the Higgs production cross section to the gluon fusion cross section
from~\cite{Dittmaier:2011ti}.

Following~\cite{Davidson:2012ds},
we require exactly one electron (assumed to come from a leptonic $\tau$ decay)
and one muon with $p_T > 30$~GeV and $|\eta| < 2.5$, and with azimuthal
separation $\Delta\phi(e, \mu) > 2.7$. The azimuthal separation between the
muon and the missing transverse momentum, $\Delta\phi(\mu, \slashed{p}_T)$ must
be less than 0.3.  The leptons are required to have opposite charge, and events
with central high-$p_T$ jets ($p_T > 30$~GeV, $|\eta| < 2.5$) are vetoed.
After this preselection, we classify events as signal-like or background-like
based on the $p_T$ of the muon and
the value of the variable
\begin{align}
  r_{\slashed{E}_T} \equiv
  \frac{\slashed{E}_T^\text{calc} - \slashed{E}_T^\text{obs}}{\slashed{E}_T^\text{obs}} \,.
\end{align}
Here, $\slashed{E}_T^\text{obs}$ is the measured missing energy and
$\slashed{E}_T^\text{calc}$ is the transverse energy of the neutrinos calculated
from the momenta of the two charged leptons under the hypothesis of a true $h \to \tau \mu$
decay. In the approximation that all decay products of the $\tau$ are collinear,
$\slashed{E}_T^\text{calc}$ is given by
\begin{align}
  \slashed{E}_T^\text{calc} =
    p_{T,e} \bigg( \frac{m_h^2}{2 E_e E_\mu (1 - \cos\theta_{e\mu})} - 1 \bigg) \,,
\end{align}
where $E_e$ and $E_\mu$ are the energies of the electron and the muon,
respectively, $p_{T,e}$ is the transverse momentum of the electron, and
$\theta_{e\mu}$ is the angle between the electron and muon momenta.

The dominant backgrounds to the search for $h \to \tau\mu$ are $Z + \text{jets}$
production with leptonic decay of the $Z$, Standard Model diboson ($WW$, $WZ$
and $ZZ$) production, single top production and $t\bar{t}$ production.
We simulate the signal and background rates in MadGraph~5
v2.0.0.beta3~\cite{Alwall:2011uj}, followed by parton showering an
hadronization in Pythia~6.426~\cite{Sjostrand:2006za}. We use the MLM
scheme~\cite{Mangano:2002ea} for matching between the matrix element and the
parton shower. For detector simulation we use PGS~\cite{PGS} with the default
implementation of the CMS detector.

\begin{figure}
  \begin{tabular}{cc}
    \includegraphics[width=0.48\textwidth]{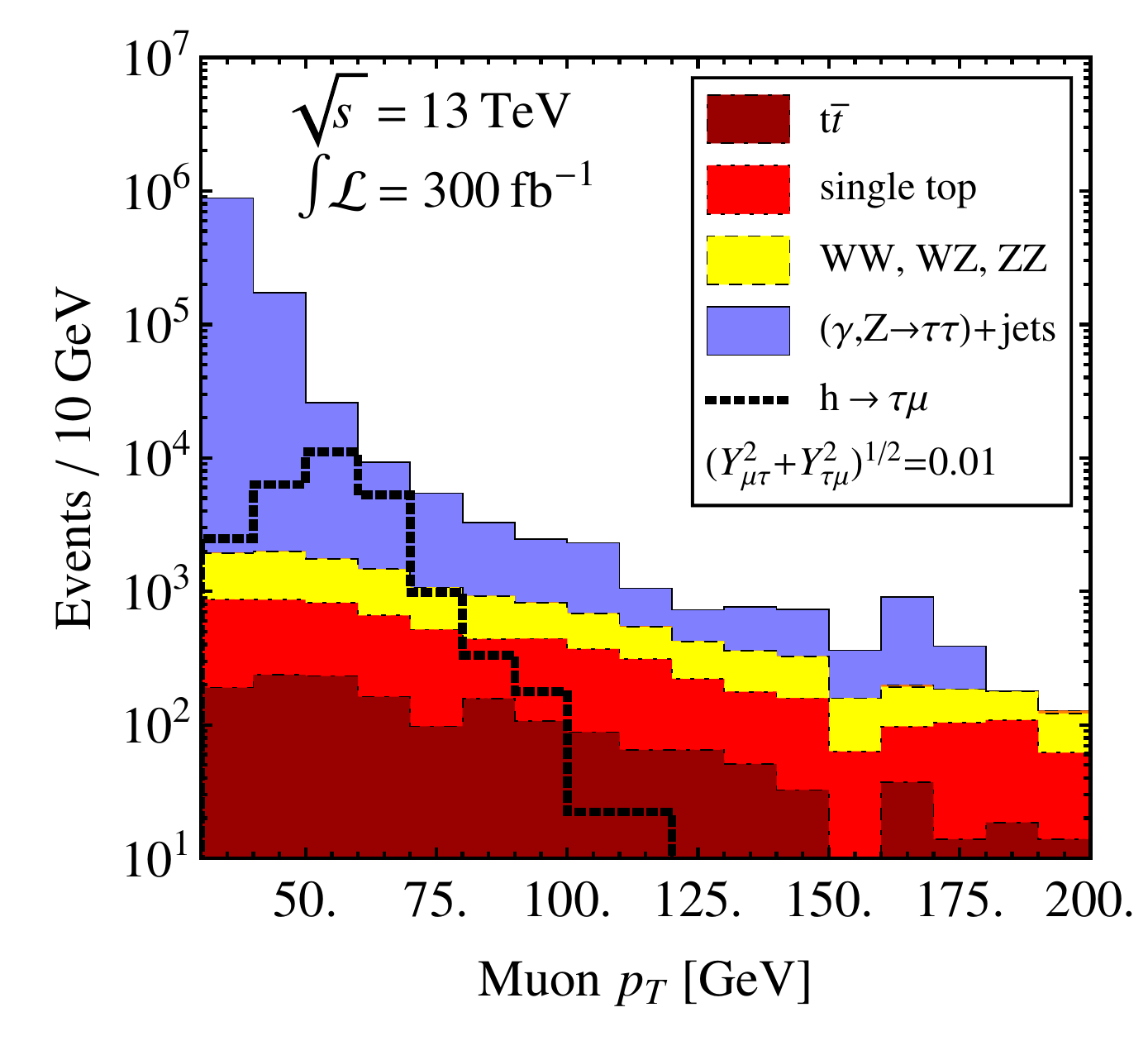} &
    \includegraphics[width=0.48\textwidth]{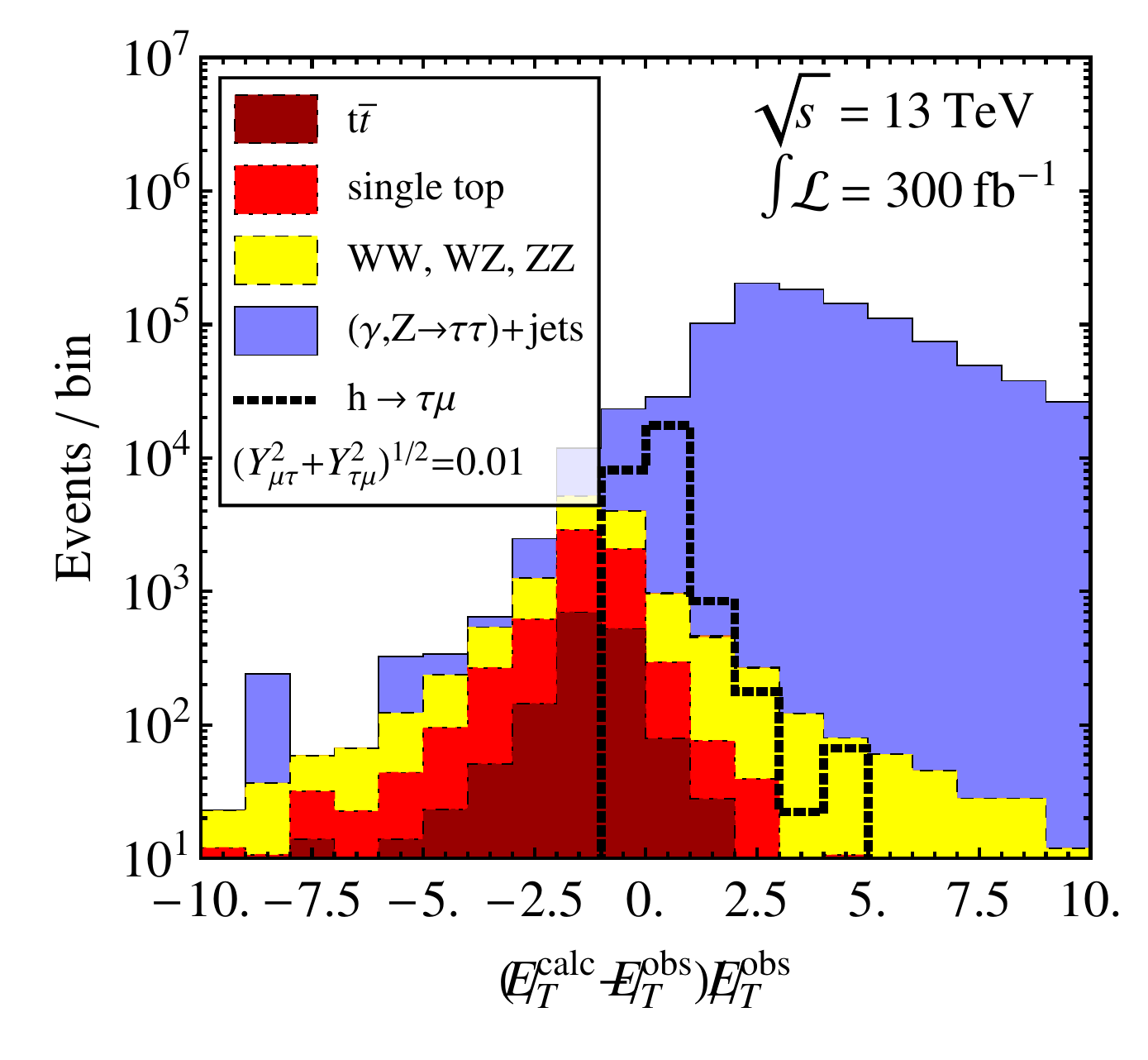} \\
    (a) & (b)
  \end{tabular}
  \caption{Distributions of (a) the $p_T$ of the muon and (b) the relative
    difference between the measured $\slashed{E}_T$ and the $\slashed{E}_T$
    calculated from the kinematics of the charged leptons.}
  \label{fig:deltaET-pTmu}
\end{figure}

The predicted distributions of $r_{\slashed{E}_T}$ and
$p_{T,\mu}$ are shown in fig.~\ref{fig:deltaET-pTmu}.  Our plots confirm
that the findings from~\cite{Davidson:2012ds} still hold at $\sqrt{s} = 13$~TeV:
the $p_T$ of the muon tends to be larger for the signal than for the
dominant $Z + \text{jets}$ background, and the difference between the measured
and calculated $\slashed{E}_T$ is typically much smaller for signal events than
for background events.

\begin{figure}
  \begin{center}
    \includegraphics[width=0.5\textwidth]{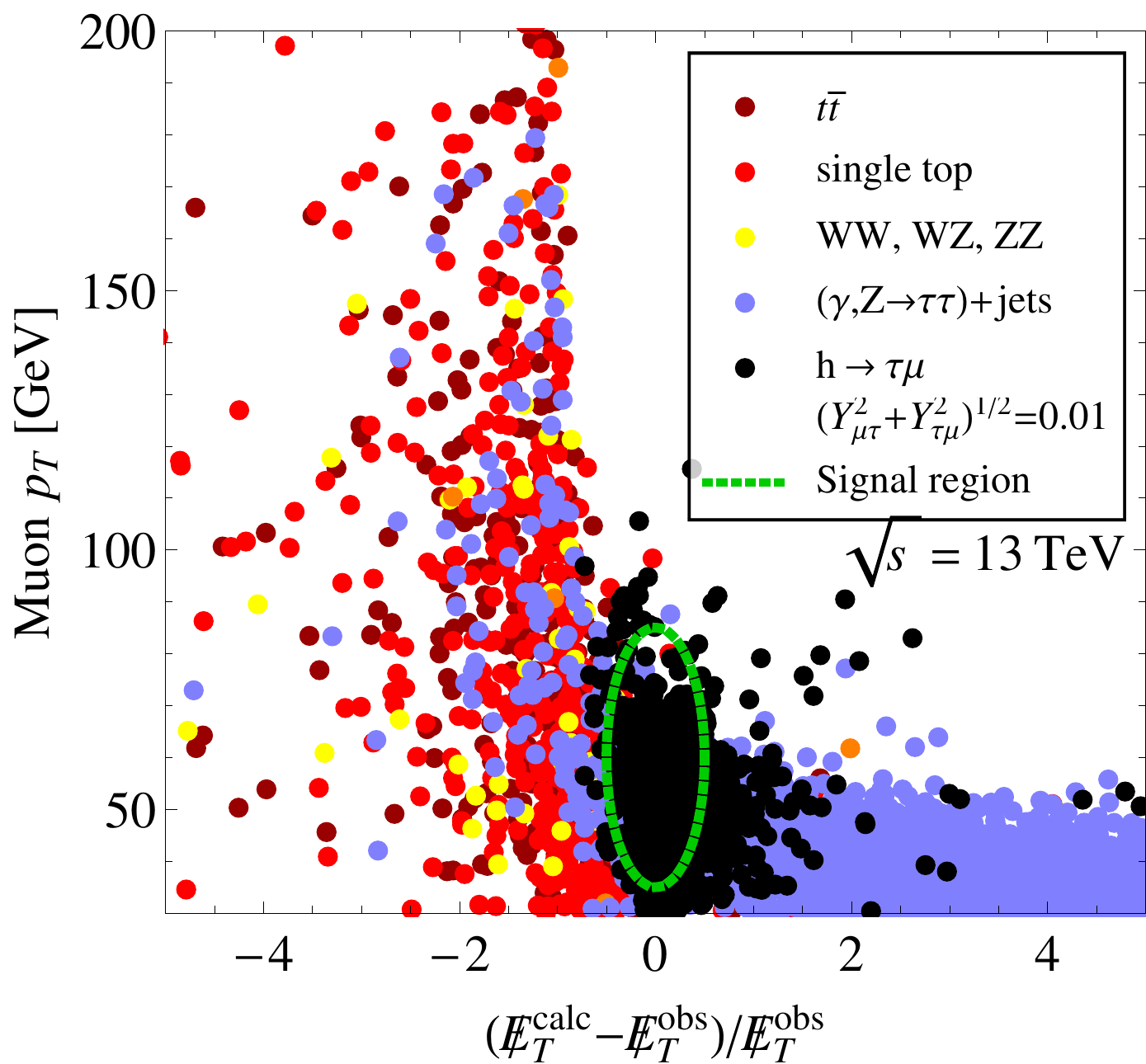}
  \end{center}
  \caption{Two-dimensional distribution of signal events (black) and background events
    (colored). The horizontal axis shows the relative difference between the
    measured $\slashed{E}_T$ and the $\slashed{E}_T$ calculated from the
    kinematics of the charged leptons.  The vertical axis shows the transverse
    momentum of the muon. The blue ellipse is the signal region defined in
    eq.~\eqref{eq:cut-ellipse}.}
  \label{fig:cuts2d}
\end{figure}

For the final selection, we require
\begin{align}
    \bigg( \frac{p_{T,\mu} - 60\ \text{GeV}}{25\ \text{GeV}} \bigg)^2
  + \bigg( \frac{r_{\slashed{E}_T}}{0.5} \bigg)^2 < 1 \,,
  \label{eq:cut-ellipse}
\end{align}
thus restricting the analysis to an ellipse in the
$p_{T,\mu}$--$r_{\slashed{E}_T}$ plane (see fig.~\ref{fig:cuts2d}).
After this final cut, the total predicted
background cross section is
\begin{align}
  \sigma_\text{BG} \simeq 64\ \text{fb} \,,
  \label{eq:sigma-bg}
\end{align}
while for the signal we obtain in the 2HDM
\begin{align}
  \sigma_\text{sig}
    \simeq 634\ \text{fb} \times \BR(h_1 \to \tau\mu)
    =       69\ \text{fb} \times
           \Bigg( \frac{\sqrt{(Y_{\tau\mu}^2 + Y_{\mu\tau}^2)
                              (\theta_{12}^2 + \theta_{13}^2)}}{0.01} \Bigg)^2 \,.
  \label{eq:sigma-sig}
\end{align}
As a crude estimate for the relative accuracy of a measurement of the rate for
$h \to \tau\mu$, we use $\sqrt{S+B} / S$, where $S$ and $B$ are the number of
signal and background events, respectively, satisfying all preselection cuts as
well as the condition \eqref{eq:cut-ellipse}. For the LHC
(300~fb$^{-1}$ integrated luminosity) and $[(Y_{\tau\mu}^2 +
Y_{\mu\tau}^2) (\theta_{12}^2 + \theta_{13}^2)]^{1/2} = 0.01$, we find
$\sqrt{S+B} / S \simeq 0.0053$.

This implies that, in the 2HDM and taking into account only statistical uncertainties, the LHC
would be able to set a 95\% confidence level upper limit $\BR(h_1 \to \tau\mu)
\lesssim 7.7 \times 10^{-4}$ or $[(\theta_{12}^2 + \theta_{13}^2)(Y_{\mu\tau}^2 +
Y_{\tau\mu}^2)]^{1/2} \lesssim 4.0 \times 10^{-4}$.
Evidence for flavor violating Higgs decays at the $3\sigma$ level would be
achievable for $\BR(h_1 \to \tau\mu)
\gtrsim 0.0013$ or $[(\theta_{12}^2 + \theta_{13}^2)(Y_{\mu\tau}^2 +
Y_{\tau\mu}^2)]^{1/2} \gtrsim 5.2 \times 10^{-4}$. Similarly $3\sigma$ evidence for CP violation
from the difference between $\BR(h_1 \to \tau^+\mu^-)$ and $\BR(h_1 \to
\tau^-\mu^+)$ requires $\BR(h_1 \to \tau\mu) \times A_{CP} \gtrsim
0.0013$.

The equivalent numbers for the effective theory from sec.~\ref{sec:EFT} are obtained
by simply setting $\theta_{12}^2 + \theta_{13}^2 = 1$ in these expressions
and replacing $Y$ by $Y^h$.

We emphasize again that the above estimates do not account for systematic
uncertainties, which would slightly decrease the sensitivity of a realistic
experimental analysis.  However, the inclusion of other decay channels---in
particular those involving hadronic tau decays and those involving same flavor
leptons---can be expected to significantly enhance the sensitivity, so
that our limits can still be considered conservative.

For the 2HDM, we show the expected 95\% C.L.\ sensitivity to CP violating
signals in $h_1 \to \tau\mu$ decays as ``Brazilian bands'' in
figs.~\ref{fig:limit-mh2} and \ref{fig:limit-theta}.  To compute these bands,
we have assumed that the observed rate for $h_1 \to \tau\mu$ decays is at the
predicted level at each parameter point, but CP is not violated in the data.
In computing the central black curve, we have therefore assumed the observed
rates for $h_1 \to \tau^+\mu^-$ and $h_1 \to \tau^-\mu^+$ to be identical.  For
parameter points below the curve, the LHC is then able to disfavor CP violation
at the actually predicted level for these parameter points at the 95\% C.L.
The green (yellow) bands are obtained in a similar way, but allowing for $1\sigma$
($2\sigma$) statistical fluctuations of the observed asymmetry away from zero.

\section{Conclusions}
\label{sec:conclusions}

To summarize, we investigated the prospects for discovering a CP asymmetry in
the flavor-violating Higgs decays $h \to \tau\mu$ and $h \to \tau e$ at the
LHC. 

Flavor violating Yukawa couplings of the SM-like 125~GeV Higgs boson appear
quite generally in models with extended electroweak symmetry breaking sectors
unless they are forbidden by the introduction of extra symmetries.  Low energy
constraints are extremely weak for couplings involving $\tau$ leptons, so that
branching ratios $\BR(h \to \tau\mu)$ and $\BR(h \to \tau e)$ of order
10\%---comparable to $\BR(h \to \tau\tau)$ in the SM---are
possible.  If the flavor violating Yukawa couplings are complex, CP violation
is possible in these decays and would manifest itself as an asymmetry between
$\BR(h \to \tau^+\mu^-)$ and $\BR(h \to \tau^-\mu^+)$ or between $\BR(h \to
\tau^+ e^-)$ and $\BR(h \to \tau^-e^+)$.

We have computed the CP asymmetries for an effective field theory with only one
Higgs boson and for a Type-III Two Higgs Doublet Model (2HDM). In the effective
theory, the asymmetries are typically suppressed by $m_\tau^2 / m_h^2$ and/or
by the Yukawa couplings $Y_{e\mu}$ and $Y_{\mu e}$, which are required to be
small due to strong constraints from $\mu \to e \gamma$. In
the 2HDM, even asymmetries of order $\text{few} \times 10\%$ are possible if one of the new Higgs
bosons is similar in mass to the SM Higgs.

We have summarized current direct and indirect constraints on flavor and CP
violating Higgs decays involving $\tau$ leptons as a function of the parameters
of the 2HDM, and we have highlighted the regions of parameter space where a
discovery of CP violation could be possible at the LHC
(see figs.~\ref{fig:limit-mh2} and \ref{fig:limit-theta}).
Interestingly, we have found that this is the case if Higgs mixing is small---a
situation that is preferred due to the so-far SM-like nature of the 125~GeV
Higgs boson.  On the other hand, CP violation at an observable level would
require that the decay $h \to \tau\mu$ or $h \to \tau e$ would have to be
observed very soon.

\section*{Acknowledgments}

We would like to thank the members of the Department of Theoretical Physics at
the Institut Jo\v{z}ef Stefan, Ljubljana, for interesting discussions and
feedback.  JK is grateful to CP$^3$ origins, where this work was initiated, and
to NORDITA Stockholm, where part of it has been carried out, for kind
hospitality.

\bibliographystyle{JHEP}
\bibliography{./higgs-cpv}

\end{document}